\documentstyle[multicol,amssymb,pra,aps]{revtex}
\newcommand{\be}{\begin{equation}}
\newcommand{\ee}{\end{equation}}
\newcommand{\bea}{\begin{eqnarray}}
\newcommand{\eea}{\end{eqnarray}}
\newcommand{\ba}{\begin{array}}
\newcommand{\ea}{\end{array}}

\begin{document}

\draft

\title{A Mechanism for Cutting Carbon Nanotubes with a Scanning Tunneling Microscope}
\author{
Angel Rubio$^{a,b}$,
S. Peter Apell$^{b,c}$,
Liesbeth C. Venema$^{d}$ and Cees Dekker$^{d}$}
\bigskip
\address{
a. Departamento de F\'{\i}sica Te\'orica, Universidad de Valladolid,
   E-47011 Valladolid, Spain\\
b. Departamento de F\'{\i}sica de Materiales, Euskal Herriko Unibertsitatea
   Aptdo. 1072 San Sebastian 20080, Basque Country \\
   and Donostia International Physics Center, San Sebastian, Spain \\
c. Department of Applied Physics, Chalmers University of Technology
   and G\"{o}teborg University, S-41296 G\"{o}teborg, Sweden \\
d. Department  of Applied Sciences and DIMES, Delft University of
   Technology, Lorentzweg 1, 2628 CJ Delft, The Netherlands }
\date{\today}
\maketitle

\widetext


\begin{abstract}
We discuss the local cutting of single-walled carbon nanotubes by
a voltage pulse to the tip of a scanning tunneling microscope. The
tip voltage ($\mid V \mid \ge $~3.8~eV) is the key physical
quantity in the cutting process. After reviewing several possible
physical mechanisms we conclude that the cutting process relies on
the weakening of the carbon-carbon bonds through a combination of
localized particle-hole excitations induced by inelastically
tunneling electrons and elastic deformation due to the electric
field between tip and sample. The carbon network releases part of
the induced mechanical stress by forming topological defects that
act as nucleation centers for the formation of dislocations that
dynamically propagate towards bond-breaking.
\end{abstract}

\pacs{PACS numbers: 61.16.Ch, 61.48.+c, 62.20.Fe}

\begin{multicols}{2}
\narrowtext

\section{Introduction}

Since the discovery of carbon nanotubes in 1991 \cite{Iijima} a
lot of progress has been made in the synthesis as well as in the
characterization of the electronic, optical and mechanical
properties of these remarkable molecules
\cite{book,book1,book2,dekker_pt}. They are promising structures
to use as components in submicrometer-scale devices
\cite{devices,pipe} and in nanocomposites \cite{nanoc}. A carbon
nanotube can be visualized as a graphite sheet rolled up
seamlessly into a cylinder. They have diameters in the range of
0.6-30~nm and are many microns in length. Depending on the
synthesis conditions they can appear in multi-walled or
single-walled configurations~\cite{Iijima,Smalley,Catherine}. The
tube symmetry determines not only the electronic (metallic or
semiconducting) character\cite{book2} but also the plastic/brittle
behavior\cite{Marco,Paul}. The special geometry makes the
nanotubes excellent candidates for mesoscopic quantum wires.
Evidence for (1D) quantum confinement was obtained from electronic
transport measurements on single-walled nanotubes~\cite{Dekker}.
The transition from one-dimensional (wire) to zero-dimensional
(quantum dot) behavior can be achieved directly by cutting a long
nanotube to a shorter length. This has been recently done by
Venema {\it et al}\cite{Venema} by applying voltage pulses to the
tip of a scanning tunneling microscope (STM) located just above a
nanotube. Discrete energy states, consistent with a 1D
particle-in-a-box model~\cite{Venema2,Rubio}, have been measured
with STM spectroscopy in such short tubes\cite{Venema2}. The
possibility to control the length of nanotubes by the cutting
technique is of interest for various applications of carbon
nanotubes in nanoscale devices~\cite{dekker_pt,devices}.

In this paper we discuss a number of possible mechanisms that can
explain the experimental observed cutting of tubes by a voltage
pulse to the STM tip for single-walled carbon nanotubes. In the
following Section the experimental data is briefly described. In
Section III, we discuss some relevant physical mechanisms that can
result in breaking of tubes and then select one as the most
promising mechanism. We analyze  the proposed cutting mechanism in
Section IV in more detail and compare the theoretical model to the
experimental results. We end the paper with a short discussion and
outlook.

\section{Experimental results}

As found earlier \cite{Venema} individual carbon nanotubes can be
locally cut by applying a voltage pulse to the tip of a scanning
tunneling microscope (STM). The carbon nanotubes that were studied
were single-walled, synthesized by a laser vaporization technique
and consisted mainly of $\sim 1.4$ nm diameter nanotubes (material
from R.E. Smalley and coworkers)\cite{Smalley}. Samples were
prepared by depositing a dispersion of nanotubes in
1,2-dichloroethane onto single-crystal Au(111) surfaces. The
experiments were done both at room temperature and at 4 K.

Nanotubes are cut by the following procedure: During imaging of a
nanotube in constant-current mode, scanning is interrupted and the
STM tip moves to a selected position on the nanotube. Feedback is
then switched off and a voltage pulse between tip and sample is
applied for 1 ms. After this pulse, the feedback is switched on
again and scanning resumes where imaging was interrupted. The
distance between the STM tip and the nanotube during a pulse is
determined by the settings for the feedback current and voltage.
Fig.~1(b) shows an example of a nanotube that has been cut into
various smaller tube pieces as a result of voltage pulses of -3.75
V applied at the positions marked in Fig. 1(a). Often, tube parts
beneath the STM tip are picked up during a pulse. This usually
leads to degradation of the tip quality. Cleaning of the tip can
then be done by applying voltages on the gold surface, away from
the nanotube.

The cutting efficiency as a function of the bias voltage applied
during a pulse is shown in Fig.~2. This experimental result
provides essential input for the theoretical modeling of the
cutting mechanism described below. The efficiency at a specific
voltage is defined as the number of successful cutting events
divided by the total number of applied pulses at that voltage. A
large number (about 150) of voltage pulses were applied at room
temperature in a range of 1 to 6 V, at positive and negative
polarity. The pulses were applied on various nanotubes, both
semiconducting and metallic. To study the dependence of cutting
efficiency on the distance between the STM tip and the nanotube
during a pulse, feedback currents were varied between 20~pA and
1~nA and feedback voltages between 0.1 and 3~V were used. No
dependence of cutting efficiency on the tunnel distance was found.
Pulses applied with various feedback currents and voltages are
therefore included in same graph of Fig.~2. The main experimental
results are listed below:

\begin{itemize}
\item  We find a sharp threshold for the voltage of 3.8 $\pm $ 0.2 V for cutting
nanotubes. This is independent of polarity. Below 3.6 V, nanotubes
could almost never be cut. Above 4 V, tubes were almost always
cut.

\item  The cutting efficiency is independent of the feedback tunnel current
or voltage. The tunnel resistance has been varied over three
orders of magnitude, which changes the tunnel distance
significantly. This demonstrates that the determining physical
quantity for cutting is the voltage, rather than the electric
field.

\item  The cutting procedure appears to be effective for different types
of nanotubes. We observe no dependence on the electronic character
(i.e. semiconducting or metallic) of a tube.

\item  Nanotubes can be cut at room temperature as well as at 4 K.

\item  Nanotubes within bundles can be cut as efficiently as isolated
single-wall tubes (see for an example Fig.~3).

\item Upon decreasing the tunnel distance considerably by increasing the tunnel
current beyond 1 nA, nanotubes are moved away laterally by the STM
tip during imaging.

\item  The separation between nanotube ends created by a cut varies
significantly in size, from a few nm to 20 nm. Sometimes, the tube
ends are displaced after the cut. Fig.~4(a) for example shows a
strongly bent nanotube on which two voltage pulses were applied
near the marked positions. Fig.~4(b) shows that the three tube
parts separated by the cuts were moved significantly by the
cutting events. Most likely, the nanotube was fixed on the
substrate under some strain that was released by the cutting.
\end{itemize}

\section{Physical concepts for the cutting mechanism}

In this Section we survey a number of possible physical mechanisms
for the breaking of nanotubes by a voltage pulse that are
interrelated to various degrees\cite{comment_binding}.


\begin{enumerate}
\item {\it Shear pressure};
Simply crashing the tip into the tube could be a possible cutting
mechanism. However, experimentally we find that the tubes are
moved laterally when the tip is brought close to the tube. This is
related to the large reversible elastic response (flexibility)
exhibited by carbon nanotubes~\cite{falvo98}. Simulations of
C$_{60}$-molecules impact on carbon nanotubes have shown that even
large radial forces produce reversible elastic distortions
indicating that crashing the tip onto the tube is not an efficient
method to cut\cite{ar1}. Furthermore, there is no obvious energy
scale of 4 eV in the crashing process.

\item  {\it Crack propagation}; This is related to the propagation of
voids and cracks already present in as-grown nanotubes under low
(tensile) loads. However, there is no evidence, from STM or
otherwise, for the presence of local defects or cracks in the SWNT
material studied here. This mechanism is hence not considered to
be important.

\item  {\it Collective excitations (plasmons)};
Plasmons excited by inelastic electron scattering of the tunneling
current  can decay into electron-hole pairs, phonons or other excitations
that induce a polarization or charge separation in the tube.
Eventually the release of the plasmon energy leads to a break
through local heating and atom evaporation. Due to the particular
cylindrical geometry of carbon nanotubes we expect to have a $\pi
$-plasmon excitation at about 5 eV~\cite{plasmon}. This has been
confirmed experimentally \cite{Pichler} for SWNTs of $\sim 1.4$ nm
diameter. The decay of the plasmon excitation into atom
evaporation is a well-known phenomenon in metallic clusters where
the surface-plasmon energy is of the order of the binding energy
leading to atom-evaporation as an effective decaying mechanism of
the collective excitations~\cite{Heer}. In the case of tubes with
an internal binding energy greater than $\sim$ 7~eV/atom (as for
most carbon solids; see Table I), however, multiple-plasmon
excitations need to be active to induce transitions which  weaken
the carbon bonds. This puts this mechanism behind first-order
models. However, it can enhance the probability of electronic
excitations (see below).

\item{\it Localized particle-hole excitations}; This mechanism
is concerned with the symmetry-allowed interband excitation of
localized $\sigma$-states to states of $\pi^*$ character near the
Fermi level. These electronic excitations leave localized
$\sigma$-holes behind which weaken the C-C bonds by creating
possible nucleation sites for breaking of the tube. Electrons
tunneling inelastically between tip and tube are the source for
these excitations. The probability for this process is favored by
the local electric field at the tip-tube interface, which,
independent of the metallic or semiconducting behavior, enhances
the number of possible bonding/antibonding transitions triggering
the bond-breaking. The density of states (DOS) for a (10,10)
nanotube, which is metallic, is shown in Fig.~5. Near the Fermi
level, the DOS is finite and constant. At higher energies, sharp
peaks can be observed which are the Van Hove singularities at the
subband onsets \cite{stm_apa}. At above/below 3.6 eV from the
Fermi level the interband excitations involve states with a
predominant $\sigma^{\ast} $/$\sigma$-localized character with a
small curvature induced $\sigma -\pi $/$\sigma ^{\ast }-\pi ^{\ast
}$ hybridization. The excitations of the $\sigma $/$\sigma ^{\ast
}$ states have been found in nanotubes to lead to a broad spectral
feature in the experimental electron-energy loss-spectra close to
the $\pi$-plasmon excitation \cite{Pichler}, similar to the case
of $\sigma -\pi ^{\ast }$ interband transitions in graphite
\cite{book1}. This interband excitation involving localized
$\sigma $-states introduces a natural energy threshold for the
cutting process to take place around 3.6~eV. This agrees quite
well  with the experimental observation of a sharp threshold
voltage for cutting of 3.8~eV that is independent of polarity.

\item  {\it Field-induced elastic deformation};
The large electric field from the tip apex causes significant
changes in the C-C bond lengths.  This introduces a mechanical
instability  in the tube that triggers the formation of
topological defects (double pentagon-heptagon defect pairs, see
Fig.~6)\cite{SW,Vin} that dynamically evolve towards breaking of
the tube (see next section for details)\cite{jump_to_contact}.
This effect is enhanced by the stress introduced in the nanotube
during the electronic excitation of localized $\sigma$-states (as
discussed in the previous mechanism). The induced stress acts on
the tube for the whole 1~ms applied voltage pulse which is long
enough for the formation and evolution of the dislocation cores.
\end{enumerate}

The combination of the last two processes is the most likely to
constitute the basic mechanism for cutting of nanotubes. The
cutting process then is triggered by the inelastic electron
excitations involving transitions of localized $\sigma$-states at
about $\sim$ 3.6 eV. This accounts for the threshold voltage found
in the experiments. The C-C bonds are further weakened by the
mechanical stress induced by the large electric field between tip
and sample. These two processes together induce topological
defects and drive the system to mechanical instability. In the
next section we present a more detailed analysis of mechanisms 4
and 5 as well as specific molecular dynamics simulations of bond
rearrangement and bond-breaking driven by the stress introduced in
the nanotube.

\section{The cutting mechanism}

In the previous section we have sorted out a possible scenario for
the cutting of nanotubes by a voltage pulse. This section
discusses in more detail the bond weakening due to mechanisms 4
and 5 described in the previous section. The formation and
propagation of topological defects due to the bond weakening will
also be discussed.

\subsection{Localized electronic excitation}

In order to understand the role of electronic excitations in the
bond-weakening\cite{dimer} and cutting process we show in Fig.~5
how the density of states of a (10,10) SWNT is modified by an
applied electric field in the direction perpendicular to the tube
axis. Here the structural relaxation and electronic calculations
were done in the framework of the ab-initio total-energy
density-functional pseudopotential theory~\cite{ar3}. We find that
the field tends to increase the nanotube's lattice parameter by a
few \%, nearly independent of the polarity of the applied bias
potential. Eventually, as the field strength increases, the
structure can reach a state which is not in a stable equilibrium
for the carbon atoms any more~\cite{Tomanek}.

In the calculations shown in Fig. 5 the field acts on the whole
tube for fixed atomic coordinates. Results are shown corresponding
to electric fields accessible in the experimental setup (up to
1~eV/\AA). One could also take into account the spatial variation
of the applied field related to the tip-size. In that case there
are distinct regions: one far from the tip where the electronic
properties of the tube are dictated by the isolated tube and the
other just beneath the tip where the electronic properties are
modified by the applied field. From Fig.~5 we see that the applied
field results in the appearance of localized levels which
increases the DOS in an energy region close to the Fermi level
(this effect is related to the bond-weakening and bond-length
increase in carbon nanotubes under an applied voltage discussed
above). The increase of density of states near the Fermi level
enhances the number of possible transitions that can take place.
Semiconducting nanotubes have a DOS comparable to that of metallic
nanotubes, but have an energy gap with zero DOS near the Fermi
energy. However, the electric field will induce states within the
gap, similar to the increase of DOS near the Fermi level for
metallic nanotubes. This allows localized particle-hole
excitations to take place also for semiconducting nanotubes.
Indeed, in the experiments no difference is found in cutting
efficiency between semiconducting and metallic nanotubes.

It may appear that any electronic mechanism should be current
dependent. However the number of broken bonds is far less than the
number of available electrons in the cutting process. The
excitations become possible because of the large current available
in the STM experiment. The experimental range of currents during
the cutting process (more than 50 nA) is such that during the 1~ms
pulse more than thousand electrons are involved in inelastic
events. This means that we are in a saturated regime where cutting
can in principle be achieved independent of the current. This
situation can be compared to the situation of cutting Si:H bonds
with an STM, where the desorption yield was found to be
independent of bias or current once the bias is large enough~
\cite{petern}. The formation of topological defects is a natural
way of releasing the excitation energy and naturally triggers the
breaking process.


\subsection{Field-induced elastic deformation}

We study the tip-field induced mechanical stress in the nanotube
within a macroscopical approach. An argument for using a
macroscopic model is that the specific elastic constants of single
wall nanotubes determined from experiments follow quite well the
predictions for the macroscopic elasticity theory (in terms of the
Young's modulus, Poisson ratio and torsion and bending elastic
constants~\cite{Hernandez}). Although the ultimate description of
the fracture mechanics is a complex phenomenon that requires both
macroscopic and microscopic descriptions, we thus rely on a
continuous model to get a first estimate of the parameters
involved in the process. The typical energy of the distortion
process is such that it corresponds to bond-breaking energies for
carbon compounds (see Table I) or to displacements which fulfill a
Lindeman melting criterion \cite{Lindemann} of a 10\% change in
bond distance.

We consider the tube as a thin hollow rod which is  clamped to the
surface by van der Waals forces. In the cutting process, the tube
is distorted by the tip in a region the size of the tip
(length-scale L). For a hollow tube (inner radius $a$ and outer
radius $b$) of bending inertia I=$\pi (b^{4}-a^{4})/4$ and Young's
modulus Y, we expect a relative deformation $\delta$ for an
applied force $F$ of \cite{LL}

\be
\frac{\delta }{L}\approx \frac{FL^{2}}{NYI} \ee
where N is a number of order 10-100 depending on details of the
modeling such as force distribution and boundary conditions. We
estimate the force F from the interaction between a sphere of
radius R (representing the tip curvature) and a tube of outer
radius b (7.5 {\AA } in our situation) to be \cite{DJ}:
\be
F=\frac{\epsilon _{0}}{d}\sqrt{Rb}(2\pi V)^{2}. \ee
where $V$ is the applied bias voltage and $d$ the tip-tube
distance. We see that it depends on the inverse separation. In
this equation we have neglected the hollow inner part of the tube
that would reduce the force by approximately 20 \%, but not its
dependence on $d$ and $V$.  With V=4 Volts, a distance of d=10
{\AA } and R=100 {\AA } (a reasonable value in STM  when
simulating a tip with a sphere \cite{Johansson}) we find a force
of 15 nN~\cite{linear}. This is independent of the polarity of the
bias, as found experimentally. Soler {\it et al.} \cite{soler}
experimentally found that graphite surface would be experiencing a
1 {\AA } deformation in an STM configuration for a force of the
order of 1 nN. Notice however that the relevant elastic constants
in their case are related to the weak interlayer bonding in
graphite as reflected in the corresponding c$_{33}$ and c$_{44}$
elastic contansts. These are much smaller than the c$_{11}$
constant (reflecting the strong intralayer $sp^2$-like bond) which
is relevant in our case (see Table II)~\cite{shear}.

Using now Eqn.~(1) and inserting Y $\sim$ 1~TPa, we find that for
a distortion $\delta /L$ higher than the 10\% Lindeman criterion,
L has at least to be about 5 nm (of course strongly depending on
N). The 10 \% change in the bond lengths can lead to brittle
behavior (see below)\cite{Marco}. All this happens for electric
fields of the order of 1 V/{\AA } which are in the experimental
range. Notice that this pull of electro-magnetic origin is
happening over the length covered by the tip shape and drops
dramatically outside the tip providing a highly ``distorted''
region.

\subsection{Dynamical bond-breaking}

Now we address how the electronically and mechanically induced
strain can lead to the breakdown of a tube. A plausible mechanism
is the Stone-Wales (SW) transformation leading to bond-rotation
defects\cite{SW}. This comprises of a pair of pentagon-heptagon
defects obtained by a simple C-C bond--rotation in the hexagonal
network, see Fig.~6. These defects are the main source of strain
release for tubes under tension~\cite{Marco,Paul} and determine
the overall electronic character of the tube \cite{Vin}. The
pentagon-heptagon defect behaves as a single edge dislocation in
the tube circumference. Once nucleated, the pair dislocations can
relax further by successive Stone-Wales transformations.

We have performed molecular dynamics simulations of strain-induced
defect and plastic/brittle behavior using the tight binding
parameterization of Ref.~\cite{tbmd} that reproduces quite well
density-functional calculations within the
local-density-approximation\cite{md}. The computed (T=0K) SW
defect formation energy in an armchair nanotube at different
applied strains shows that for a tensile strain of about 5\% for a
(10,10) tube and 12\% for a (12,0) tube, the defect geometry is
energetically favourable over the perfect tube\cite{defect_paul}.
The activation barrier for the formation of defects is lowered by
the applied tension. Indeed the activation barrier for bond
rotation in a (10,10) nanotube is found to reduce from 5.6 eV at
zero strain to 3 eV at 10\% strain; similar reduction is observed
for the barrier for separation of the pentagon-heptagon
dislocation cores. The dynamics of these defects are dictated by
the tube chirality as well as the applied tension and temperature.
In particular at high applied strain and low temperature all tubes
are brittle~\cite{Marco}. In the simulations restricted to
armchair tubes, we observed that after the nucleation of the first
SW-defect, octagonal and higher order rings start to appear  as
the applied strain is increased. This leads to brittle behavior
and shows the important role of the strain induced by the applied
voltage. The simulation time scale for the formation and evolution
of these defects are of the order of nanoseconds, much shorter
than the applied voltage pulse of 1 ms. We thus expect the natural
nucleation and evolution of these defects beneath the tip region
of the STM, leading to a bond breaking and possible local collapse
of the structure.

\section{Conclusions and Outlook}

From the previous discussion we conclude that the applied voltage
at the STM tip creates: (i) excitation of localized
$\sigma$-bonding states, (ii) a change in the density of states
around the Fermi level making the tubes more metallic-like and,
(iii) an overall distortion of the region beneath the tip, leading
to the formation and evolution of strain-induced topological
defects. These processes together are responsible for the cutting
mechanism with the threshold voltage of $\sim 4$~V. The proposed
mechanism relies only on the applied voltage. It appears that
within the experimental values, the tip-tube distance does not
change the results.

It would be interesting to attempt the cutting in multi-walled
carbon nanotubes also. The cutting will probably not work for
these nanotubes since the inner layers are able to accomodate the
stress acting on the outer layers. Furthermore we expect the
inter-tube van der Waals interaction in multi-walled nanotubes to
be strong enough  to partially  release the concentration of
elastic strain so that only atoms of the outer surface will be
affected.  This is different to the case of nanotubes in bundles,
where the cutting process is as effective as for isolated
nanotubes (see Fig. 3). In this case the other tubes in the bundle
act as a global support similar to the gold substrate for the
isolated SWNTs.

It may be interesting to see if the breaking process is
accompanied by photon emission to any significant degree to gain
further information about the energetic of the possible processes.
Photon emission provides a signal that is dependent on the local
physical environment beneath the tip with a spatial resolution
determined by the size of the local creening charge (colective
mode) involved in the photon emission. Typical resolution can be
estimated as $\sqrt{Rd} \sim 20${\AA} using our length scales
introduced above~\cite {Johansson,comment_field}. A related topic
would be field emission from nano-tubes.

To summarize, we have characterized in detail the possible
mechanisms responsible for the breaking up of tubes as found
experimentally. Nanotube cutting is a promising technique for the
emerging field of nano-manipulation and nanodevices. More studies
will have to be made to understand all the interesting physics
taking place at this nanoscale level.

\section*{Acknowledgments}

\indent This work was supported in part by grants from the Swedish
Natural Science Research Council, Iberdrola S.A, JCyL (Grant:
VA28/99) and the European Community TMR contracts
ERB-FMRX-CT98-0198 and ERBFMRX-CT96-0067 (DG12-MIHT). SPA and AR
are grateful to the enlightening atmosphere at the University of
the Basque Country where most of this work was done. We thank
H.L.J. Temminck Tuinstra for help with the experiments and
acknowledge discussions with J.W.G. Wild\"oer, D. Tomanek, P.
Bernier and M. Buongiorno-Nardelli. The work at Delft was
supported by the Dutch Foundation for Fundamental Research of
Matter (FOM).
%


\setbox2 = \hbox{\cite{Hernandez}}
\begin{table}
\caption{ Typical spring constants k [N/m], energy D$_{o}$ [eV]
needed to break a bond at T=0, and equilibrium distances R$_{eq}$
({\AA}) for different carbon bonds. In rough terms
displacing an atom by 0.1 {\AA} corresponds to an elastic energy
change of 5-10 eV.
}
\begin{tabular}{c|c|c|c|c|}
& bond & k (N/m)$^*$ & D$_{o}$ (eV)$^*$ & R$_{eq}$ ({\AA})$^*$ \\ \hline
& C-H & 460 & 4.0 & 1.11 \\ & C-C & 440 & 3.4 & 1.52 \\
 & C=C &960 & 7.4 & 1.34 \\ & C$\equiv$C & 1560 & 10.0 & 1.21 \\
\end{tabular}
$^*$ M.F. Ashby, Acta. Metall. {\bf 37}, 1273 (1989).
\end{table}

\setbox3 = \hbox{\cite{book1}}
\begin{table}
\caption{Typical elasticity parameters Y (Young's modulus) and
$\sigma_{Y}$ (yield stress) for different carbon materials\box3~
compared to steel. Due to the large anisotropy in the elastic
constants of graphite we add those in parenthesis. They are
relevant for the elastic strength and flexibility of the single
wall nanotubes (SWNT).}
\begin{tabular}{c|c|c|c|}
& Material & Y (GPa) & $\sigma_{Y}$ (GPa) \\ \hline
& Steel & 200 & 3 \\
& Fullerenes & 400 & $\simeq$ 12 \\
& Diamond & 1140 & 50-90 \\
& & (c$_{11}$=1090; c$_{12}$=120; c$_{44}$=640) &  \\
& Graphite & 700 & $\simeq$ 40 \\
&  & (c$_{11}$=1060; c$_{33}$=36.5; c$_{44}$=4.5) &  \\
& Tubes & SWNT 1250 $^a$ &  \\
&  & SWNT-ropes $^b$ $\sim$ 1000 & $\sim$ 1 $^b$ \\
&  & (TB calc. Y=1240; $\nu$=0.26)\cite{Hernandez} &  \\
\end{tabular}
$^a$ A.~Krishnan, E.~Dujardin, T.W.~Ebbesen,
P.N.~Yanilos and M.M.J.~Treacy, Phys. Rev. B {\bf 58}, 14013 (1998).\\
$^b$J.P. Salvetat, G.A.D. Briggs, J.M. Bonard, R.R. Bacsa,
A.J. Kulik, T. St\"{o}ckli, N.A. Burnham and L. Forro, Phys. Rev.
Lett. {\bf 82}, 944 (1999)
\end{table}


\begin{figure}
\caption[]{ Room temperature images of a carbon nanotube (a)
before and (b) after it was cut at three positions marked with
crosses in (a).}
\end{figure}

\begin{figure}
\caption[]{Cutting efficiency versus applied voltage. Efficiency
is here defined as the number of successful cuts divided by the
total number of applied pulses at that specific voltage. About 150
voltage pulses were applied on various nanotubes. The feedback
current was varied between 20~pA and 1~nA; the feedback voltage
between 0.1 V and 3V.}
\end{figure}

\begin{figure}
\caption[]{ Room temperature image of a bundle of nanotubes. An
individual nanotube within the bundle, indicated by the arrow,
could be cut by a voltage pulse.}
\end{figure}

\begin{figure}
\caption[]{Room temperature images of a strongly bent nanotube (a)
before and (b) after it was cut near the two positions marked with
crosses in (a). The three separated tube parts appear to be
displaced as a result of the cutting events. Image (b) is an
enlargement of the area within the square in (a).}
\end{figure}

\begin{figure}
\caption[]{Density of states of an armchair (10,10) single-wall
nanotube for various applied electric fields perpendicular to the
tube axis. There is an increase of the density of states at the
Fermi level that leads to bond weakening (see text). We also show
the results for a field of 4 eV/\AA\ as the dashed line in the
bottom panel. For comparison the density of states for a single
graphene layer (dash line) is also given in the top panel.}
\end{figure}

\begin{figure}
\caption[]{Schematic description of the strain-induced topological
pentagon-heptagon pair defect (Stone-Wales
transformation)~\cite{SW}. The tube axis is in the vertical
direction. }
\end{figure}

\end{multicols}


\begin{thebibliography}{99}


\bibitem{Iijima}  S. Iijima, Nature {\bf 354}, 56 (1991); P.M. Ajayan and S.
Iijima, Nature {\bf 361}, 333 (1993).

\bibitem{book}  See for example the special issues on {\it Nanotubes} in
Carbon {\bf 33} (1996); J. Appl. Phys. A {\bf 67} (1998); {\bf 68} (1999).

\bibitem{book1}  M.S. Dresselhaus, G. Dresselhaus, P.C. Eklund, {\it Science
of Fullerenes and Carbon Nanotubes} (Academic Press Inc., San Diego, 1996);
M.S. Dresselhaus, G. Dresselhaus, K. Sugihara, I.L. Spain and H.A. Goldberg,
{\it Graphite Fibers and Filaments}, Springer Verlag (1988).

\bibitem{book2}  T.W. Ebbesen, {\it Carbon nanotubes: preparation and
properties} (CRC Press, New York, 1997).

\bibitem{dekker_pt} C. Dekker, Physics Today {\bf 52}, 22 (1999).

\bibitem{devices}
 S.J. Tans, A.R.M. Verschueren and C.Dekker, {\ Nature}, {\bf 393} 49 (1998).


\bibitem{pipe} J. Liu, A. G. Rinzler, H. Dai, J. H. Hafner, R. K. Bradley, P.
J. Boul, A. Lu, T. Iverson, K. Shelimov, C. B. Huffman, F.
Rodriguez-Macias, Y. Shon, T. R. Lee, D. T. Colbert, and R. E.
Smalley, Science, {\bf 280}, 1253 (1998). P.G. Collins, A. Zettl,
H. Bando, A. Thess, R.E. Smalley, Science {\bf 278}, 100 (1997).

\bibitem{nanoc} S. Curran, P.M. Ajayan, W. Blau, D.L.
Carroll, J.. Coleman, A.B. Dalton, A.P.Davey, A. Drudy, B. McCarthy, S.
Maier and  A. Strevens,  Advanced Materials 10, 1091 (1998);
 P.M. Ajayan, L.S. Schadler, C. Giannaris and A. Rubio
 (submitted for publication).

\bibitem{Smalley}  A. Thess, R. Lee, P. Nikolaev, H. Dai, P. Petit, J.
Robert, C. Xu, Y.H. Lee, S.G. Kim, A.G. Rinzler, D.T. Colbert, G.E.
Scuseria, D. Tomanek, J.E. Fisher, R.E. Smalley, Science {\bf 273}, 483
(1996)

\bibitem{Catherine}  C. Journet, W. Maser, P. Bernier, A. Loiseau, P.
Deniard, S. Lefrant, R. Lee, J. Fischer, Nature {\bf 388}, 756 (1997).



\bibitem{Marco}  M. Buongiorno-Nardelli, B.I. Yakobson and J. Bernhold,
Phys. Rev. Lett. {\bf 81}, 4656 (1998); Phys. Rev. B {\bf 57}, 4277 (1998).

\bibitem{Paul}  P. Zhang, P.E. Lammert and V.H. Crespi, Phys. Rev. Lett.
{\bf 81}, 5346 (1998).

\bibitem{Dekker}  S.J. Tans, M.H. Devoret, H. Dai, A. Thess, R.E. Smalley,
L.J. Geerligs, C. Dekker, {\ Nature} {\bf 386}, 474 (1997); M.
Bockrath, D.H. Cobden, P.L. McEuen, N.G. Chopra, A. Zettl, A.
Thess and R.E. Smalley, {\ Science} {\bf 275}, 1922 (1997).

\bibitem{Venema}  L.C. Venema, J.W.G. Wild\"{o}er, H.J.L. Temminck Tuinstra,
C. Dekker, A.G. Rinzler and R.E. Smalley, Appl. Phys. Lett. {\bf 71}, 2629
(1997).

\bibitem{Venema2}  L.C. Venema, J.W.G. Wild\"{o}er, S.J. Tans, J.W. Janssen,
H.L.J. Temminck Tuinstra, L.P. Kouwenhoven and C. Dekker, Science, {\bf 283}%
, 52 (1999).

\bibitem{Rubio}  A. Rubio, D. Sanchez-Portal, E. Artacho, P. Ordej\'{o}n and
J.M.Soler, Phys. Rev. Lett. {\bf 82}, 3520 (1999).


\bibitem{comment_binding} As a starting point for all the possible
cutting mechanisms a strong binding between the tube and the
substrate is necessary either chemically or through van der Waals
forces. Experiments [T. Hertel, R. Martel and P. Avouris, J. Phys.
Chem. {\bf 102}, 910 (1998); Phys. Rev. B {\bf 58} 13870 (1998)]
and first-principle calculations~\cite{Rubio} show that tubes have
a fairly good binding to the gold substrate. It is typically of
the order of 2~meV/atom and adds up to a significant amount of
binding energy for a tube.


\bibitem{falvo98}S.~Iijima, C.~Brabec, A.~Maiti and J.~Bernholc, J. Chem. Phys.
{\bf 104} 2089 (1996); M.R.~Falvo, G.J.~Clary, R.M.~Taylor,
V.~Chi, F.P.~Brooks, S.~Washburn and R.~Superfine, Nature {\bf
389}, 582 (1997).


\bibitem{ar1} A. Lordi and N. Yao, J. Chem. Phys. {\bf 109}, 2509
(1998).

\bibitem {plasmon} Note that the
imaginary part of the inverse dielectric function for graphite
(sheet) has a peak around 6-7 eV, related to a collective plasma
excitation~\cite{book1}. Geometric arguments indicate that for a
cylinder this excitation should move down in energy by $\sqrt{2}$.

\bibitem{Pichler}  T. Pichler, M. Knupfer, M.S. Golden, J. Fink, A. Rinzler
and R.E. Smalley, Phys. Rev. Lett. {\bf 80}, 4729 (1998).

\bibitem{Heer}  W.A. de Heer, Rev. Mod. Phys. {\bf 65}, 611 (1993);
M. Brack, Rev. Mod. Phys. {\bf 65}, 677 (1993).

\bibitem{stm_apa} A. Rubio, Appl. Phys. A {\bf 68} 275 (1999).
Although interactions
with the metallic substrate or  other tubes introduces some modifications
in the shape of the van Hove singularities and the opening of
pseudogaps, the main
electronic structure is well described by the isolated tube.

\bibitem{SW} A.J. Stone and D.J. Wales, Chem. Phys. Lett. {\bf 128},
501 (1986).

\bibitem{Vin}  V.H. Crespi, M.L. Cohen and A. Rubio, Phys. Rev. Lett.
{\bf 79}, 2093 (1997)

\bibitem{jump_to_contact} The mechanical deformation might also lead,
in some conditions, to jump to contact of tip and sample with
potential material transfer between the two. The jump-to-contact
voltage can be estimated from the derivative of the elastic and
electrostatic forces in equilibrium [O. Hansen, J.T. Ravnkilde, U.
Quaade, K. Stokbro and F. Grey, Phys. Rev. Lett. {\bf 81}, 5572
(1998)], to be of the order of a few volts depending on the
effective tip-sample distance. The jump-to-contact voltage is very
sensitive to the equilibrium tunnel gap. This indicates that it is
unlikely that we are in this regime here.

\bibitem{dimer} Removing or adding an electron from/to a tube
bond can introduce stresses in the network. So, for instance, the
bond length of the C$_{2}$-dimer increases from 1.24 to 1.30 {\AA}
when taking one electron out, which is a 5\% effect. However in
more confined structures (e.g. benzene) the effect is less
relevant as long as the electron is removed from the $\pi$-cloud.

\bibitem{ar3} W.E. Pickett, Comput. Phys. Rep. {\bf 9}, 115 (1989);
 M.C. Payne, M.P. Teter, D.C. Allan, T.A. Arias, J.D. Joannopoulos,
 Rev. Mod. Phys. {\bf 64}, 1045 (1992).


\bibitem{Tomanek}
The calculations show a slight asymmetry  in the response of the
carbon nanotube with respect of the applied bias polarity for
potentials larger than $\pm 4$eV (in agreement with calculations
performed by D. Tomanek, private communication 1999). Note that
the field introduces a quite large mechanical stress in the
system.  We can estimate from a simple displaced harmonic
oscillator model (displacement $\Delta x$) the field strength
k$\Delta x$/e necessary for a 10\% change in bond distance
(Lindemann melting criterion\cite{Lindemann}). Using k-values from
Table I we obtain values of the order of 4V/{\AA} for the single
bond. This corresponds to a change in elastic energy of 6-7 eV.
Once this energy is higher than bonding-antibonding transition
energies it cannot be accommodated and the bond will tend to
break.


\bibitem{petern} Ph. Avouris, R.E. Walkup, A.R Rossi, H.C. Akpati, P.
Nordlander, T.-C. Shen, G.C. Abeln and J.W. Lyding, Surf. Sci.
{\bf 363}, 368 (1996).

\bibitem{Hernandez}  E. Hern\'{a}ndez, C. Goze, P. Bernier and A. Rubio,
Phys. Rev. Lett. {\bf 80}, 4502 (1998); L. Vaccarini, C. Goze, L.
Henrard, E. Hern\'{a}ndez, P. Bernier and A. Rubio, Carbon  (in press).


\bibitem{Lindemann}  F.A. Lindemann, Z. Physik {\bf 11}, 609 (1910).


\bibitem{LL}  L. D. Landau and E. M. Lifschitz, {\em Theory of Elasticity},
2nd ed. (Pergamon, Oxford 1975).

\bibitem{DJ}  B. Derjagin, Kolloid Z. {\bf 69}, 155 (1934);
L.R. White, J. Colloid interface Sci. {\bf 95}, 286 (1983).
To compute the force we use the
Derjagin approximation, that was later on elaborated by
White. In this approach, the force between two
objects is related to the curvatures involved and the interaction
energy $E(d)$ per unit area for the corresponding {\em planar}
situation: $F= 2\pi R_{eff}E(d)$
where $d$ is the separation between the two solids (average distance
between tip and tube in our situation) and R$_{eff}$ expresses the
curvatures involved and is given by:
$ R_{eff}=R\sqrt{\frac{b}{R+b}}\approx \sqrt{Rb}$.
The latter approximation is valid when R$\gg $b, which applies to
our case. The interaction energy for two planar surfaces making up
a capacitor with a bias voltage  V is the classical charging
energy: $E(d)=\frac{1}{2}CV^{2}$.
For C=$4\pi \epsilon _{0}/d$ (per unit area) we then get the final
result for the force given in Eqn.(2).

\bibitem{Johansson}  P. Johansson, R. Monreal and P. Apell, Phys. Rev. B{\bf %
42}, 9210 (1990).

\bibitem{linear} It is known from other works
(H.F. Budd and J. Vannimenus, Phys. Rev. Lett. {\bf 31}, 1218
(1973); Phys. Rev. B{\bf 12}, 509 (1975);
S. Andersson, B.N.J. Persson, M. Persson and N.D. Lang,
Phys. Rev. Lett. {\bf 52}, 2073 (1984))
that the classical
force used by invoking Eq.(2) actually reflects an average over the true
force acting. There is also a first order
response which we should consider since our system is truly microscopic.
Within linear response the total force F acting on the ions  from an
external field E has to average to zero. We checked that
this linear term contribution is very small in our particular case.


\bibitem{soler}  J.M. Soler, A.M. Baro, N. Garc\'\i a and H. Rohrer, Phys. Rev.
Lett. {\bf 57}, 444 (1986).

\bibitem{shear} The Young's modulus, Y=k/r$_{o}$ where r$_{o}$
is an interatomic distance, will be of the order of TPa (see Table
II). As the shear stress $\mu =Y/(1+\nu )$ (where $\nu \approx
0.26$ is the Poisson~\cite{Hernandez}) is of the same order of magnitude as
the Young modulus therefore, we concentrate all the discussion of
the cutting on the strain.

\bibitem{tbmd}  C. H. Xu, C. Z. Wang, C. T. Chan, and K. M. Ho, J. Phys.
Condens. Matter {\bf 4}, 6047 (1992).
\bibitem{md} More exhaustive calculations have
been done in refs.~\cite{Marco,Paul} using a semiempirical Tersoff
parametrization of the carbon-carbon interaction.
\bibitem{defect_paul}Above this critical value the defect density
increases rapidly with length~\cite{Paul}.

\bibitem{comment_field} Higher resolution and photon intensities can
be achieved by taking advantage of the optical-field enhancement
at the STM-tip when the tip-surface cavity is illuminated on with
a laser. The field enhancement factor is of the order of 1000 for
typical STM-tips and substrates as highly oriented pyrolytic
graphite or gold (A.V. Bragas, S.M. Landi and O.E. Mart\'{\i }nez,
Appl. Phys. Lett. {\bf 72}, 2075 (1998)). This new optical
microscopy can also be used to study the linear and nonlinear
optical response of carbon nanotubes and other nanostructures.

\end{thebibliography}
\end{document}